\documentclass[twocolumn,showpacs,preprintnumbers,amsmath,amssymb,superscriptaddress]{revtex4}

\usepackage{graphicx}
\usepackage{dcolumn}
\usepackage{bm}

\newfont{\boldit}{cmbxti10}
\newfont{\boldslant}{cmbxsl10}
\def\op{\psi}

\def\FEss{{\cal F}_{\rm s}}
\def\FEem{{\cal F}_{\rm e}}

\def\FErel{{\cal F}_{\rm rel}}
\def\FEeff{{\cal F}_{\rm eff}}
\def\FErig{{\cal F}_{\rm rig}}
\def\FErigtil{{\widetilde{\cal F}}_{\rm rig}}
\def\op{\psi}
\def\tc{T_{\rm c}}
\def\c{c_{\alpha\beta}}
\def\u{u_{\alpha\beta}}
\def\aab{a^{(1)}_{\alpha\beta}}
\def\rs{\rho^{\rm S}}
\def\smhalf{{\scriptstyle\frac{\scriptstyle 1}{\scriptstyle 2}}}
\def\smquar{{\scriptstyle\frac{\scriptstyle 1}{\scriptstyle 4!}}}
\def\klb{k_{\rm B}}

\begin{document}

\title{
Squeezing superfluid from a stone:\\
Coupling superfluidity and
elasticity in a supersolid}

\author{Alan T.~Dorsey}
\affiliation{Department of Physics,
University of Florida, P.O.~Box 118440, Gainesville, FL 32611-8440}

\author{Paul M.~Goldbart}
\affiliation{Department of Physics,
University of Illinois at Urbana-Champaign,
1110 West Green Street, Urbana, IL 61801}

\author{John Toner}
\affiliation{Institute for
Theoretical Science, Department of Physics, University of Oregon,
Eugene, OR 97403}

\date{\today}

\begin{abstract}
Starting from the assumption that the normal solid to supersolid
(NS-SS) phase transition is continuous, we develop a
phenomenological Landau theory of the transition in which
superfluidity is coupled to the elasticity of the crystalline $^4$He
lattice. We find that the elasticity does not affect the universal
properties of the superfluid transition, so that in an unstressed
crystal the well-known $\lambda$-anomaly in the heat capacity of the
superfluid transition should also appear at the NS-SS transition. We
also find that the onset of supersolidity leads to anomalies in the
elastic moduli and thermal expansion coefficients near the
transition and, conversely, that
inhomogeneous lattice strains can
induce local variations of the superfluid transition temperature,
leading to a broadened transition.
\end{abstract}


\maketitle

Superfluidity---the ability of {\it liquid\/} $^4$He, when cooled
below 2.176~K, to flow without resistance \cite{kapitza38,allen38}
through narrow pores---has long served as a paradigm for the
phenomenon of ``off-diagonal long-range order" (ODLRO) in quantum
liquids and superconductors \cite{yang62}.
Supersolidity---the coexistence of ODLRO with the crystalline order
of a solid---was proposed theoretically
\cite{andreev69,chester70,leggett70,matsuda70,
fernandez74,saslow77,liu78,pomeau94,bijlsma97} as an even more
exotic phase of {\it solid\/} $^4$He, but it has eluded detection
\cite{bishop81,meisel92}.
Recently, Kim and Chan \cite{kim_nature_2004,kim_science_2004} have
reported the onset of ``nonclassical rotational inertia"
\cite{leggett70} in a torsional oscillator experiment with solid
$^4$He, and they interpret their results as indicating the onset of
supersolidity.
However, their interpretation remains controversial
\cite{ceperley04,saslow05,prokofev05,burovski05, galli05,dash05}, so
it is important to complement the nonequilibrium torsional
oscillator measurements with {\it equilibrium\/} thermodynamic
measurements, e.g., of the specific heat. In this work we start from
the assumption that normal solid to supersolid (NS-SS) phase
transition is continuous, and develop a phenomenological Landau
theory of the transition in which superfluidity is coupled to the
elasticity of the crystalline $^4$He lattice.
We find that the elasticity does not affect the universal properties
of the superfluid transition, so that in an unstressed crystal the
well-known $\lambda$-anomaly in the heat capacity of the superfluid
transition should also appear at the NS-SS transition. We also find
that the onset of supersolidity leads to anomalies in the elastic
moduli and thermal expansion coefficients near the transition;
conversely, inhomogeneous strains in the lattice can induce local
variations of the superfluid transition temperature, leading to a
broadened transition.
As our theory is rooted in a few simple assumptions and
symmetry principles, we expect our results to be robust and
insensitive to the details of a microscopic model for the
supersolidity.

We hypothesize that, as with the super{\it fluid\/} $^4$He, the
appropriate order parameter describing the onset of supersolidity is
a complex scalar field $\op({\bf r})$, depending on the
location ${\bf r}$. We shall assume that the phase transition to the
supersolid state is continuous, as is the super{\it fluid\/} $^4$He
transition.  Then, as is well known from the theory of critical
phenomena~\cite{Ma}, the {\it universal\/} properties of the
supersolid transition may be obtained via a model that retains only
those terms in the free energy up to leading (relevant) order in
powers of the fields and their spatial gradients, resulting in the
Landau form:
\begin{equation}
\FEss\!\!=\!\!
\int \!\! d^{3}\! r
\left\{
\smhalf
\c\,\partial_{\alpha}\op\,\partial_{\beta}\op^{\ast}\!\!+
\smhalf a(T)
\vert\op\vert^{2}\!\!+
\smquar w
\vert\op\vert^{4}
\right\}.
\label{LGWPSI}
\end{equation}
Here, $a(T)$ depends smoothly on the temperature $T$; it is negative
at low temperatures, changing sign slightly above the transition
temperature $T_{\rm c}$. [If fluctuations of $\op$ are ignored,
$T_{\rm c}$ would be exactly the temperature at which $a(T)$ changes
sign.]\thinspace\ For $T<T_{\rm c}$ ($T>T_{\rm c}$) the thermal expectation
value of $\langle\op\rangle$ is non-zero (zero).   The constant $w$
measures the strength of the nonlinearity, and the symmetric tensor
$c_{\alpha\beta}$ characterizes the spatial anisotropy inherited
from the crystallinity of the normal solid. For an isotropic
superfluid or for a cubic crystal $c_{\alpha\beta} =
c\,\delta_{\alpha\beta}$, with $c$ a constant, whereas for an hcp
crystal (such as solid helium \cite{SolidHe}) $c_{\alpha\beta}$ is
uniaxial, such that $c_{\alpha\beta}= c_z n_\alpha n_\beta + c_\perp
(\delta_{\alpha\beta} - n_\alpha n_\beta)$, with ${\bf n}$ a unit
vector that points along the preferred axis of the crystal, and
$c_z$ and $c_\perp$ independent constants.
It is important to note that the symmetry of the superfluid density
tensor $\rho^{\rm S}_{\alpha\beta}$ (which relates the superfluid
velocity to the momentum density) is the same as that of
$c_{\alpha\beta}$.

If we were dealing with the normal-to-super{\it fluid\/} transition,
Eq.~(\ref{LGWPSI}) (with $c_{\alpha\beta}=c\,\delta_{\alpha\beta}$)
would be the entire story. However, because we are dealing with the
normal- to super{\it solid\/} transition, it is not: when
formulating theories of continuous phase transitions, it is
necessary to keep {\it all\/} of the degrees of freedom that are
soft (i.e.~exhibit large thermal fluctuations) at the transition
\cite{Ma}. For a super{\it fluid\/}, $\op$ {\it is\/} the only such
degree of freedom.  However, in a super{\it solid\/} there are
additional phonon degrees of freedom that are soft not only at the
transition but throughout both the normal and supersolid phases.
These are associated with displacements ${\bf u}({\bf r})$ of the
positions ${\bf r}$ of the undistorted crystal lattice. (The
normal-solidification of the $^4$He occurs at a temperature
substantially higher than $T_{\rm c}$ so that amplitude fluctuations
of the density waves are not soft and may be neglected.)

As the free energy must be invariant under spatial translations and
rotations~\cite{landau_elasticity}, it can only depend upon
${\bf u}({\bf r})$ through the symmetric strain tensor
$u_{\alpha\beta}\equiv\frac{1}{2} \left(
\partial_{\alpha}u_{\beta}+
\partial_{\beta}u_{\alpha}+
\partial_{\alpha}u_{\gamma}\,\partial_{\beta}u_{\gamma}
\right)$. Thus, at long wavelengths the relevant terms in the free
energy involving $\u$ alone are simply those of standard elastic
theory: $\FEem=\frac{1}{2}\int d^{3}r\,
\lambda_{\alpha\beta\gamma\delta}\,u_{\alpha\beta}\,u_{\gamma\delta}$,
where $\lambda_{\alpha\beta\gamma\delta}$ are the bare elastic
constants, and repeated indices are summed over. The form of
$\lambda_{\alpha\beta\gamma\delta}$ is dictated by the symmetry of
the crystal; for an hcp crystal such as $^4$He, it is parameterized
by 5 independent elastic constants~\cite{landau_elasticity}.

To determine the form of the coupling between the supersolid order
parameter $\op$ and the local displacement field ${\bf u}({\bf r})$,
we follow the work of Aronovitz \textit{et al}\rlap.
\cite{aronovitz90} by allowing the (formerly constant) parameters in
$\FEss$ [i.e.~$a(T)$ and $w$] to depend on the local value of ${\bf
u}$ in a manner consistent with the symmetries of the system.  Thus,
we expand in powers of the strain tensor,
$a(T)\rightarrow a^{(0)}+
a_{\alpha\beta}^{(1)}\,u^{\phantom{(1)}}_{\alpha\beta}+
a_{\alpha\beta\gamma\delta}^{(2)}\, u^{\phantom{(1)}}_{\alpha\beta}\,
u^{\phantom{(1)}}_{\gamma\delta}+\cdots, \label{EQ:Texpansion}$
and similarly for
$w$.  Here, the tensor $a_{\alpha\beta}$ has the same symmetry as
$\c$, and hence for a uniaxial crystal (such as the hcp phase of
solid helium \cite{SolidHe}) $a^{(1)}_{\alpha\beta}= a_z n_\alpha
n_\beta + a_\perp (\delta_{\alpha\beta} - n_\alpha n_\beta)$, with
$a_z$ and $a_\perp$ independent coupling constants.

The terms of $O(u^{2}_{\alpha\beta})$ in the expansion of $a$
and of $O(u_{\alpha\beta})$
in the expansion of $w$ all prove, by
na{\"\i}ve power counting~\cite{Ma}, to be irrelevant; i.e., they do
{\it not\/} affect the universal critical properties of the
transition.  Hence, these properties follow
from
the following minimal model:
\begin{eqnarray}
\FErel&=& \int d^{3}r\ \left\{
     \smhalf
     c_{\alpha\beta}\,
     \partial_{\alpha}\op\,\partial_{\beta}\op^{\ast}
     +
     \smhalf a^{(0)}
     \vert\op\vert^{2}
     +
     \smquar w
     \vert\op\vert^{4}
     \right.
\nonumber\\
      &&\quad\qquad
      \left.
      +
      \smhalf
      \lambda_{\alpha\beta\gamma\delta}
      \,u_{\alpha\beta}\,u_{\gamma\delta}
      +
      \smhalf
      a_{\alpha\beta}^{(1)}\,u_{\alpha\beta}
      \vert\op\vert^{2}
      \right\}.
\label{EQ:minimal}
\end{eqnarray}
\textit{In our minimal model the effect of the elasticity is to
produce a local, strain-dependent critical temperature for the
superfluid.} In fact, our minimal model is formally equivalent to
that of a magnetic system
of planar spins [an $O(2)$ ferromagnet] on a
compressible lattice, for which locally dilating or compressing the
lattice causes the exchange couplings to decrease or increase,
resulting in a local change of the critical temperature. With this
analogy in hand, we may use the work of de Moura
\textit{et al}.~\cite{demoura76},
which shows that the elasticity is irrelevant to
the critical properties of the $O(2)$ ferromagnet, provided the
specific-heat exponent $\alpha$ of the decoupled
(i.e.~$a_{\alpha\beta}^{(1)}=0$) system is negative, which it is at
$d=3$ \cite{Liposuction03}. Thus, we may conclude that the
universality class, and hence universal properties, associated with
the supersolid transition are unaltered by the coupling of elastic
degrees of freedom to the supersolid order parameter.  In
particular, this implies a $\lambda$-anomaly in the specific heat
$C_p$ near the transition:
\begin{equation}
C_p(t)= \frac{A^{\pm}}{\alpha} |t|^{-\alpha} (1 + a_c^{\pm}
|t|^\Delta
       + b_c^{\pm}|t|^{2\Delta} + \cdots) + B,
\label{alpha}
\end{equation}
where $t\equiv(T-T_{\rm c})/T_{\rm c}$ denotes the reduced
temperature, $\alpha=-0.0127\pm 0.0003$ is the universal specific
heat exponent of the superfluid transition, and $\Delta=0.529\pm
0.009$ is the equally universal correction to scaling exponent
\cite{Liposuction03}. The subscripts $+$ and $-$ in
Eq.~(\ref{alpha}) denote behaviors above ($T > T_{\rm c}$) and below
($T < T_{\rm c}$) the transition, respectively. Although the
constants $A_\pm$ are non-universal (they will, e.g., change as one
moves along the SS-NS phase boundary in the pressure-temperature
phase diagram), their {\it ratio is\/} universal; the current best
estimate of its value is $A_{+}/A_{-}= 1.053\pm 0.002$
\cite{Liposuction03}. The superfluid density tensor
$\rs_{\alpha\beta}$ also exhibits universal scaling with reduced
temperature $t$, i.e.,
$\rs_{\alpha\beta}={}^{0}\!\rs_{\alpha\beta}\,|t|^\nu$, where the
tensor $^{0}\!\rs_{\alpha\beta}$ is temperature independent, and
$\nu = 0.67155\pm0.00027$ is \cite{Liposuction03} the universal
correlation-length exponent.

Despite the irrelevance of the elastic couplings to $\op$ for the
universality class of the transition, these couplings {\it do\/}
have important, experimentally observable consequences: due to
them, the elastic properties inherit singularities in their
temperature dependence from parent singularities associated with the
critical fluctuations of the supersolid order parameter.
   Following Ref.~\cite{aronovitz90},
we construct the
effective free energy $\FEeff$ governing the elastic fluctuations
by integrating out the supersolid fluctuations:
\begin{equation}
e^{-\FEeff[u_{\alpha\beta}]/\klb T}\equiv
\int {\cal D}(\op,\op^{\ast})
e^{-\FErel[\psi,u_{\alpha\beta}]/\klb T}.
\end{equation}
Proceeding perturbatively, in powers of the $\psi$-$u_{\alpha\beta}$
coupling term, we obtain (neglecting an additive constant)
\begin{equation}
\!\!\FEeff[u_{\alpha\beta}]\!=\!\! \int\! d^{3}r \left\{
\!\frac{1}{2}(\lambda_{\alpha\beta\gamma\delta}\!+
\delta\lambda_{\alpha\beta\gamma\delta}) u_{\alpha\beta}
u_{\gamma\delta}\!+ \sigma_{\alpha\beta}u_{\alpha\beta }\right\}\!,
\label{EQ:renorm-elast}
\end{equation}
where the singular parts of the fluctuation corrections to the bare
elastic constant and stress tensors are given, to leading order, by
\begin{subequations}
\begin{eqnarray}
\delta\lambda_{\alpha\beta\gamma\delta} &=&
-\,a_{\alpha\beta}^{(1)}\,a_{\gamma\delta}^{(1)}\,\widehat{C}(t)/\klb
T,
\\
\sigma_{\alpha\beta}
&=&
+\,a_{\alpha\beta}^{(1)}\,D(t).
\label{thermexpan}
\end{eqnarray}
\end{subequations}
Here, the governing functions, $\widehat{C}$ (which proves to be
proportional to the singular part of the specific heat) and $D$, are
given by
\begin{subequations}
\begin{eqnarray}
D(t)
\!\!&\equiv&\!\!\!
\left\langle
\frac{\vert\op({\bf r})\vert^{2}}{2}
\right\rangle_{0} =
\frac{\partial\FErigtil}{\partial a^{(0)}},
\\
\widehat{C}(t) \!\!&\equiv&\!\!\! \lim_{k\to 0}\!\int\!\!\!
\frac{d^{d}k^{\prime}}{(2\pi)^{d}} \left\langle\!
\frac{\vert\op\vert_{\bf k}^{2}}{2} \frac{\vert\op\vert_{{\bf
k}^{\prime}}^{2}}{2} \!\right\rangle_{0}^{\rm c} \!\!=\! -\klb
T\frac{\partial2 \FErigtil}{\partial {a^{(0)}}^2},
\end{eqnarray}%
\label{dellambda}%
\end{subequations}%
where the expectation values $\langle\cdots\rangle_{0}$ are taken
with respect to the {\it rigid\/} $\op$ system (i.e.~a system in
which all $u_{\alpha\beta}$'s are frozen at zero), whose free energy
$\FErig$ is just
$\FEeff[u_{\alpha\beta}]$ with all
$u_{\alpha\beta}$'s set to zero.
Moreover, $\vert\op\vert_{\bf k}^{2}$
denotes the Fourier transform of $\vert\op({\bf r})\vert^{2}$, and
$\langle\cdots\rangle^{\rm c}$ indicates a connected correlator.  In
addition, $\FErigtil$ is the free energy density associated
with $\FErig$.

$D(t)$ and $\widehat{C}(t)$ can readily be related to the specific
heat singularity of the rigid system, by noting that $a^{(0)}$ is a
linear function of $T$ close to $\tc$, so that derivatives with
respect to $a^{(0)}$ are proportional (near $T_{\rm c}$) to
derivatives with respect to $T$.  Hence, simple thermodynamic
identities imply that $D(t)$ and $\widehat{C}(t)$ are proportional
to the entropy and the specific heat of the rigid system,
respectively, near $T_{\rm c}$.  This immediately determines  their
critical behavior: $D(t)=G^\pm |t|^{1-\alpha}(1 + a_c^\pm
|t|^\Delta+\cdots)$ and $\widehat{C}(t)=A'^\pm |t|^{-\alpha}(1 +
a_c^\pm |t|^\Delta+\cdots)$, where $G^+/G^- =A'^+/A'^- = A^+/A^-=
1.053\pm 0.002$ are universal.

It is evidently valuable to estimate the size of the expected
elastic and thermal expansion anomalies.  To do this, we need an
estimate of the couplings $\aab$ in Eq.~(\ref{EQ:minimal}), and this
can be obtained,
   following Ref.~\cite{bergman76},
from the form of the SS-NS phase boundary, $T_{\rm c}(P)$, in the
pressure-temperature phase diagram. In particular, the couplings are
related to the slope of the boundary via
\begin{eqnarray}
{\partial \tc (P) / \partial P} \sim
a^{(1)}/\left(\lambda a^{\prime}\right),
\label{slope}
\end{eqnarray}
where $a^\prime\equiv da^{(0)}/dT$, and $\lambda$ is a typical
elastic constant. With this in hand, we can now estimate the size of
the elastic and thermal expansion anomalies. This can be
accomplished by using Eq.~(\ref{dellambda}) to estimate
$\widehat{C}$ and $D$ well away from the critical point (say, at
$T=2\tc$), where we can make the Gaussian approximation to the
correlation functions, which gives, e.g.,
\begin{equation}
D(T)\vert_{T=2\tc}\sim \int d^{3}k \frac{k_B T}{a^{(0)}(T)} \sim
\frac{k_B}{a^{\prime}\,\xi_{0}^{3}},
\end{equation}
where $\xi_{0}$ is the high-temperature correlation length for $\op$
fluctuations. We have taken the integral over $k$ to have an
ultraviolet cutoff comparable to $\xi_{0}$, used the fact that, well
above $\tc$, $a^{(0)}(T)/c\sim\xi_{0}^{-2}$ to replace the
propagator [up to $O(1)$ factors] by $a^{(0)}$ for all $k$, and
estimated $a(T=2\tc)\sim a^{\prime}\tc$. For super{\it fluid\/}
$^4$He, the length $\xi_0$ is known to be comparable to the atomic
size: $\xi_{0}\sim 0.2\,{\rm nm}$; for want of better information,
we shall assume that this is also true for the supersolid.

By using this estimate for $D(T)$ in Eq.~(\ref{thermexpan}),
and then minimizing Eq.~(\ref{EQ:renorm-elast}) over $\u$,
we arrive at a typical value for the thermal expansion:
\begin{equation}
\delta\u \sim \frac{a^{(1)}}{\lambda}D(T) \sim
\frac{\partial}{\partial P} \frac{\klb\tc(P)}{\xi_{0}^{3}},
\end{equation}
where in the last step we have used Eq.~(\ref{slope}). Estimating
$|\partial\tc (P)/\partial P|\sim 10^{-2}\ {\rm K/atm}$ by its value
in the {\it liquid\/} state of $^4$He~\cite{SolidHe} gives a
typical value of $\delta\u\sim 0.16$. For the fractional anomaly in
the elastic constants, arguments essentially identical to those just
used give
\begin{equation}
\frac{\delta\lambda}{\lambda}\sim \frac{\lambda
\klb}{\tc\xi_{0}^{3}} \left(\frac{\partial \tc (P)}{\partial
P}\right)^{2}\sim 0.17.
\end{equation}

Perhaps the best way to observe the predicted anomalies in the
elastic constants is through sound speed measurements. In {\it
single crystals\/}, the polarization of the sound modes studied must
be chosen judiciously: the uniaxial form for the expansion
coefficients, $a^{(1)}_{\alpha\beta}= a_z n_\alpha n_\beta + a_\perp
(\delta_{\alpha\beta} - n_\alpha n_\beta)$, implies that only
``bulk'' elastic moduli, specifically, the terms
$\lambda_{zz}
(n_\alpha n_\beta u_{\alpha\beta})^{2}+
\lambda_{\perp \perp}
((\delta_{\alpha\beta}-  n_\alpha n_\beta) u_{\alpha\beta})^{2}+
\lambda_{\perp z}
((\delta_{\alpha\beta}-  n_\alpha n_\beta)
u_{\alpha\beta})(n_\alpha n_\beta u_{\alpha\beta})$,
acquire anomalous temperature dependence. The elastic constants
$\lambda_{zz}$, $\lambda_{\perp \perp}$, and $\lambda_{\perp z}$ are
readily shown by a standard sound mode analysis~\cite{dig it} to
only affect the sound speeds of modes polarized in the plane formed
by the normal to the hexagonal layers and the direction of
propagation.  Furthermore, in this plane, transverse modes
propagating either along, or orthogonal to, the layers also have
sound speeds independent of the anomaly-displaying elastic constants
$\lambda_{zz}$, $\lambda_{\perp \perp}$, and $\lambda_{\perp z}$.
Hence, to observe the anomaly in the sound speeds in a single
crystal of supersolid hcp helium, one should study modes polarized
in the plane formed by the hexagonal layers and the direction of
propagation, {\it and\/} choose that propagation direction
{\it not\/} to lie in, or orthogonal to, the layers.

It is, however, unlikely that experiments will be performed on
single crystals of helium.  It is far more likely that they will be
performed on polycrystalline samples, which are macroscopically
isotropic, due to the random orientations of the constituent
crystallites. Calculating the isotropic shear modulus $\mu$ and
Lam\'{e} coefficient $\lambda$ of such an ensemble of randomly
oriented crystallites is well-known to be a formidable problem (see,
e.g., Ref.~\cite{KF}). Nonetheless, using the exact bounds of
Hill~\cite{fool on the}, we can show~\cite{dig it} that {\it both\/}
the shear modulus and the bulk modulus of a macroscopically
isotropic polycrystalline helium sample will exhibit the
$|t|^{-\alpha}$ anomaly we predict here. As a result, {\it both\/}
the transverse {\it and\/} the longitudinal sound speeds of such a
sample will show the $|t|^{-\alpha}$ anomaly.

Beyond the critical properties described above, our model has the
important implication that in a helium crystal the SS-NS transition
would be {\it rounded\/} by any spatially inhomogeneous internal
stresses that make $a^{(1)}_{\alpha\beta}\, u^{\phantom{(1)}}_{\alpha\beta} \neq 0$.
As such stresses are almost unavoidable in any crystal (and are
believed to be present in the experiments of Kim and
Chan~\cite{kim_nature_2004,kim_science_2004}), this rounding is
almost certain to be present in all experiments performed to date.
The reason for such broadening is very simple: stresses that make
$a_{\alpha\beta}^{(1)}\,
 u^{\phantom{(1)}}_{\alpha\beta}$ spatially inhomogeneous
make $T_{\rm c}$ of the SS-NS transition spatially inhomogeneous as
well; cf.~Eq.~(\ref{EQ:minimal}). Therefore, roughly speaking,
distinct parts of the sample would become supersolid at distinct
temperatures, broadening the transition.  Such broadening is evident
in the $\rs (T)$ vs.~$T$ plots of Kim and
Chan~\cite{kim_science_2004}, which clearly {\it do not\/} show the
expected $\vert{t}\vert^{\nu}$ singularity near the putative $T_{\rm
c}$, with $\nu$ being the correlation length exponent. [If they did,
$\rs (T)$ vs.~$T$ would hit the horizontal axis
{\it perpendicularly\/}, rather than---as in those
data---tangentially.]\thinspace\ This broadening may also explain
the apparent absence of the expected $\lambda$-anomaly in the
specific heat~\cite{clark05} near the putative $T_{\rm c}$ in those
experiments: this peak is simply ``smeared away\rlap.\rq\rq

Our model  suggests that the best way to definitively observe (or
rule out) the critical behavior of the specific heat, elastic
constants, superfluid density, etc.~that we predict here, and
thereby to test the notion that helium actually {\it does\/} exhibit
a supersolid phase, is to prepare the solid helium samples in a way
that is designed very carefully to eliminate any spatially
inhomogeneous stresses in the crystal.  At the very least, this
would require that samples always be pressurized at high
temperatures in the liquid phase, and only then cooled into the
solid phase; this cooling should be done at {\it constant\/}
pressure, so as to avoid introducing inhomogeneous strains.

One intriguing, albeit highly speculative, final implication of the
coupling of strains to the supersolid order parameter $\op$ has to
do with the very existence of the supersolid state in $^{4}$He.
A number of microscopic calculations \cite{ceperley04} suggest that
$^{4}$He does {\it not\/} have a supersolid state at all, in
contradiction with the experiments of
Chan~\textit{et al}.~\cite{kim_nature_2004}.
These calculations were presumably done under conditions of purely
{\it hydrostatic\/} stress (i.e.~simple pressure), for which only
the three diagonal components of the strain tensor are non-zero and
equal (i.e.~$u_{xx}=u_{yy}=u_{zz}\neq0$).
If it were the case that the coefficients $a_z$ and $a_\perp$ in the
uniaxial coupling tensor for solid hcp $^4$He happened to obey
$a_z\approx -2 a_\perp$ then the effect of this pressure on the
effective $\tc$ for supersolid order would be very small.  If the
{\it unstrained\/} $\tc$ were negative [i.e.~the coefficient
$a^{(0)}$ in Eq.~(\ref{EQ:minimal}) were positive for all $T$], this
would imply that, under such a hydrostatic stress, the crystal would
never enter the supersolid state.  If, however, the crystal were to
be subjected to an {\it anisotropic\/} stress (i.e.~one for which
the relation $u_{xx}=u_{yy} =u_{zz}$ is {\it not\/} satisfied), the
near cancellation of
$a^{(1)}_{\alpha\beta}\,u^{\phantom{(1)}}_{\alpha\beta}$
would {\it not\/}
occur, and this term might be able to make the effective $\tc$
positive [i.e.~change the overall sign of the coefficient of
$|\op|^2$ in Eq.~(\ref{EQ:minimal})]. That is, it is possible in our
model that, although a {\it hydrostatically\/} stressed sample would
{\it not\/} show a supersolid phase, an {\it anisotropically\/}
stressed one might. This could be highly significant, as Chan
\textit{et al}.~\cite{kim_nature_2004}~believe that such
anisotropic stresses {\it are\/} present in their samples.
Although the above argument is obviously quite speculative, it seems
to us, nonetheless, a possibility that these random stresses, far
from being an experimental nuisance, might just be what is
responsible for the presence of supersolidity.  Support for this
idea comes from two other facts: (1)~Many experiments
\cite{bishop81,meisel92} do not see supersolidity. Perhaps these
samples simply lacked sufficiently large inhomogeneous, anisotropic
stresses. (2)~The superfluid fraction in the experiments of Chan
\textit{et al}.~\cite{kim_nature_2004} is extremely low.  Could
this be because only very small, highly anisotropically stressed
regions of the sample are going supersolid?

In summary, by assuming that the NS-SS transition is
continuous and using general symmetry arguments, we
have constructed a minimal model for understanding
the NS-SS transition. Our model predicts a rich
phenomenology, whose existence, if observed, would
help corroborate the torsional oscillator evidence
for the supersolid phase.

The authors gratefully acknowledge valuable discussions with D.\
Ceperley, M.\ W.\ Meisel, and C.-D.\ Yoo, and the support of the
Aspen Center for Physics, where this work was initiated.  This work
was supported in part by the U.S.~Department of Energy (P.M.G.).

\end{document}